\documentclass[12pt,preprint]{aastex}
\shorttitle{TYPE II BURSTS AND CMES}
\shortauthors{Krupar et al.} 

\begin{document}


\title{AN ANALYSIS OF INTERPLANETARY SOLAR RADIO EMISSIONS ASSOCIATED WITH A CORONAL MASS EJECTION}


\author{V. Krupar\altaffilmark{1,2}}
\affil{The Blackett Laboratory, Imperial College London, London, UK}
\affil{Institute of Atmospheric Physics CAS, Prague, Czech Republic}
\email{v.krupar@imperial.ac.uk}

\author{J. P. Eastwood\altaffilmark{1}}
\affil{The Blackett Laboratory, Imperial College London, London, UK}
\email{jonathan.eastwood@imperial.ac.uk}


\author{O. Kruparova\altaffilmark{2}}
\affil{Institute of Atmospheric Physics CAS, Prague, Czech Republic}
\email{ok@ufa.cas.cz}

\author{O. Santolik\altaffilmark{2,3}}
\affil{Institute of Atmospheric Physics CAS, Prague, Czech Republic}
\affil{Faculty of Mathematics and Physics, Charles University, Prague, Czech Republic}
\email{os@ufa.cas.cz}

\author{J. Soucek\altaffilmark{2}}
\affil{Institute of Atmospheric Physics CAS, Prague, Czech Republic}
\email{soucek@ufa.cas.cz}

\author{J. Magdaleni\'c\altaffilmark{4}}
\affil{Solar-Terrestrial Center of Excellence-SIDC, Royal Observatory of Belgium, Brussels, Belgium}
\email{jasmina.magdalenic@oma.be}

\author{A. Vourlidas\altaffilmark{5}}
\affil{The Johns Hopkins University Applied Physics Laboratory, Laurel, MD, USA}
\email{angelos.vourlidas@jhuapl.edu} 

\author{M. Maksimovic\altaffilmark{6}}
\affil{LESIA, UMR CNRS 8109, Observatoire de Paris, Meudon 92195, France}
\email{milan.maksimovic@obspm.fr}

\author{X. Bonnin\altaffilmark{6}}
\affil{LESIA, UMR CNRS 8109, Observatoire de Paris, Meudon 92195, France}
\email{xavier.bonnin@obspm.fr}

\author{V. Bothmer\altaffilmark{7}}
\affil{Institut f\"ur Astrophysik, G\"ottingen University, G\"ottingen, Germany}
\email{bothmer@astro.physik.uni-goettingen.de}

\author{N. Mrotzek\altaffilmark{7}}
\affil{Institut f\"ur Astrophysik, G\"ottingen University, G\"ottingen, Germany}
\email{nmrotzek@astro.physik.uni-goettingen.de}

\author{A. Pluta\altaffilmark{7}}
\affil{Institut f\"ur Astrophysik, G\"ottingen University, G\"ottingen, Germany}
\email{apluta@astro.physik.uni-goettingen.de}

\author{D. Barnes\altaffilmark{8}}
\affil{RAL Space, Rutherford Appleton Laboratory, Harwell Campus, UK}
\email{david.barnes@stfc.ac.uk}

\author{J. A. Davies\altaffilmark{8}}
\affil{RAL Space, Rutherford Appleton Laboratory, Harwell Campus, UK}
\email{jackie.davies@stfc.ac.uk}


\author{J. C. Mart\'inez Oliveros\altaffilmark{9}}
\affil{Space Sciences Laboratory, University of California, CA, USA}
\email{oliveros@ssl.berkeley.edu}

\author{S. D. Bale\altaffilmark{9,10}}
\affil{Space Sciences Laboratory, University of California, CA, USA}
\affil{Physics Department, University of California, CA, USA}
\email{bale@ssl.berkeley.edu}


\altaffiltext{1}{The Blackett Laboratory, Imperial College London, London, UK}
\altaffiltext{2}{Institute of Atmospheric Physics CAS, Prague, Czech Republic}
\altaffiltext{3}{Faculty of Mathematics and Physics, Charles University, Prague, Czech Republic}
\altaffiltext{4}{Solar-Terrestrial Center of Excellence-SIDC, Royal Observatory of Belgium, Brussels, Belgium}
\altaffiltext{5}{The Johns Hopkins University Applied Physics Laboratory, Laurel, MD, USA}
\altaffiltext{6}{LESIA, UMR CNRS 8109, Observatoire de Paris, Meudon, France}
\altaffiltext{7}{Institut f\"ur Astrophysik, G\"ottingen University, G\"ottingen, Germany}
\altaffiltext{8}{RAL Space, Rutherford Appleton Laboratory, Harwell Campus, UK}
\altaffiltext{9}{Space Sciences Laboratory, University of California, CA, USA}
\altaffiltext{10}{Physics Department, University of California, CA, USA}

\begin{abstract}
Coronal mass ejections (CMEs) are large-scale eruptions of magnetized plasma that may cause severe
geomagnetic storms if Earth-directed.
Here we report a rare instance with comprehensive in situ and remote sensing observations of a CME
combining white-light, radio, and plasma measurements from four different vantage points.
For the first time, we have successfully applied a radio direction-finding technique to an interplanetary type II burst detected by two identical widely separated radio receivers.
The derived locations of the type II and type III bursts are in general agreement with the white light CME reconstruction. We find that the radio emission arises from the flanks of the CME, and are most likely associated with the CME-driven shock. Our work demonstrates the complementarity between radio triangulation and 3D reconstruction techniques for space weather applications.

\end{abstract}


\section{INTRODUCTION}
Type II and type III bursts are generally associated with solar eruptive events \citep{1950AuSRA...3..387W,1950AuSRA...3..541W,1958SvA.....2..653G,1980SSRv...26....3M}.
Both are generated, via the plasma emission mechanism, when beams of suprathermal electrons interact with ambient plasma generating
radio emissions at the plasma frequency $f_{\rm{p}}$ (the fundamental emission)
or its second harmonic $2f_{\rm{p}}$ (the harmonic emission).
As the electron beams propagate outward from the Sun, radio emissions are generated at progressively lower frequencies
corresponding to a decreasing ambient density.
Type II bursts are generated by electron beams accelerated at the shock fronts ahead of propagating CMEs,
while type III bursts are a consequence of impulsively accelerated electrons associated with solar flares
\citep[e.g.][]{1998GeoRL..25.2493R,2001SoPh..204..121R,2000GeoRL..27..145G,2003SSRv..107...27C,2007SpWea...5.8001C,2015A&A...580A.137K}.
Although type II bursts generated in the solar corona are routinely measured from the ground,
spacecraft observations of type II bursts originating further from the Sun, in the interplanetary medium -- especially with a good signal to noise ratio --
are relatively rare \citep{1998JGR...10329651R,2005JGRA..11012S07G,2004ASSL..314..223V}.
Moreover, interplanetary type II bursts are usually patchy and intermittent with short
periods of radio enhancements \citep{1998GeoRL..25.2501K,2003ApJ...590..533R}.
It has been suggested that these enhancements are related to CME--CME or CME--streamer interactions \citep{2001ApJ...548L..91G,2012JGRA..117.4105X}.

Over the last decade, we have benefited from multipoint radio measurements obtained
by the twin-spacecraft Solar TErrestrial RElations Observatory (\textit{STEREO}) mission \citep{2008SSRv..136..487B,2008SSRv..136..529B}, launched in 2006,
which allow us to localize radio sources in the interplanetary medium between $2.5$~kHz and $16$~MHz 
\citep[e.g.][]{2009SoPh..tmp..100R,2010ApJ...720.1395T,2012SoPh..279..153M,2015SoPh..290..891M}.
Using direction-finding radio data recorded by the \textit{STEREO-B} and \textit{Wind} spacecraft,
\citet{2012ApJ...748...66M} present an analysis of an interplanetary type II burst associated with a CME--CME interaction.
More recently, \citet{2014ApJ...791..115M} performed a direction-finding analysis of another event using the same pair of spacecraft
suggesting that an interplanetary type II burst was, in this case, related to a CME-streamer interaction.

Here, we use data obtained by
the \textit{STEREO}/Waves/High Frequency Receiver (125~kHz~--~2~MHz),
which allows us to localise radio sources with distances from $5$~solar radii ($1~\rm{R_S}=695,500$~km) above
the Sun's surface up to $1$~astronomical unit \citep[1~au$=149,598,000$~km;][]{2008SSRv..136..549C,2010AIPC.1216..284K,2012JGRA..11706101K}.
We also analyze data from the \textit{STEREO}/Sun Earth Connection Coronal and Heliospheric Investigation \citep[SECCHI;][]{2008SSRv..136...67H} package, the MErcury Surface, Space ENvironment, GEochemistry, and Ranging (\textit{MESSENGER}) spacecraft and by the Solar Heliospheric Observatory (\textit{SOHO}) \citep{2007SSRv..131....3S,1995SoPh..162....1D}.

In this letter, we present a study of CME kinematics using white-light, radio and plasma measurements obtained by the \textit{STEREO}, \textit{SOHO},
and \textit{MESSENGER} spacecraft. We demonstrate that interplanetary solar radio emissions can be used for estimating both CME speeds and directions.

\section{OBSERVATIONS AND ANALYSIS}
      \label{S-data}

The CME of particular interest first appeared in the field of view of the \textit{SOHO}/LASCO C2 coronagraph
at 17:24~UT on 2013 November 29.
At the time of the event, the \textit{STEREO-A} and \textit{STEREO-B} spacecraft were $150^\circ$
of heliocentric Earth equatorial (HEEQ) longitude ahead of and $147^\circ$ behind the Earth, at heliocentric distances of $0.96$~au and $1.08$~au, respectively.
\textit{SOHO} was at the First Sun-Earth Lagrangian point (L$_1$), located some 1.5 million km upstream of the Earth.
\textit{MESSENGER} was at a HEEQ longitude of $132^\circ$ and a heliocentric distance of $0.40$~au.

\subsection{White-light Observations}

This favorable configuration of SOHO and STEREO allows us to apply the graduated cylindrical shell \citep[GCS;][]{2009SoPh..256..111T} model to coronagraph images simultaneously
recorded by these three spacecraft in order to reconstruct the CME leading edge in three dimensions (3D).
We performed this analysis using images taken between 20:00 UT and 23:30 UT on November 29, when the CME was well observed by all three spacecraft. 
From GCS fitting, we estimate the radial distance of the leading edge $r_{\rm{GCS}}$,
leading to a calculated average CME speed of $761 \pm 13$~km~s$^{-1}$
with a launch time of 19:42~UT.  The CME angular half-width $\lambda_{\rm{GCS}}$ was estimated to be $59^\circ$.
The HEEQ longitude of propagation of the CME from GCS fitting was $128^\circ$, which is roughly towards \textit{MESSENGER}.
To calculate the speed further out in the heliosphere, we also applied the self-similar expansion fitting (SSEF) technique to
\textit{STEREO-A}/HI data \citep[][Equation 6]{2012ApJ...750...23D}.
Using $\lambda_{\rm{GCS}}$ as input, SSEF yielded a CME radial speed of $862 \pm 8$~km~s$^{-1}$, liftoff at 19:17~UT, and a HEEQ propagation longitude of $135^\circ$.
These values are in very good agreement with those from the GCS fit.
However, note that the GCS and SSEF predicted liftoff times are two hours later than when the CME was first seen by \textit{SOHO},
as the CME accelerated considerably near the Sun's surface \citep{2015ApJ...806....8G}.
Inspection of the EUVI images shows that the liftoff time is about 16:00~UT
and the onset can be characterized as streamer blowout CME.

\subsection{Radio Emission}

This CME was accompanied by interplanetary type II and type III bursts detected by both \textit{STEREO} spacecraft on
November 29 -- 30, 2013 (Figure \ref{fig:spectra}).
Since the CME was launched from the far side of the Sun, ground-based high frequency radio measurements of the associated radio bursts are absent.
Figures \ref{fig:spectra}a and \ref{fig:spectra}b display the radio flux density $S$ measured by \textit{STEREO-A} and \textit{STEREO-B}, respectively.
The type III burst at 22:00~UT  on November 29 is followed by the intermittent type II burst, over an interval of seven hours.
Within this, we were able to identify three intervals when the signal at both spacecraft was intense enough
to enable successful direction-finding analysis:
at 22:30~UT (1,500~kHz) on November 29 and at 00:30~UT (600~kHz) and 04:00~UT (300~kHz) on the next day.

A recent \citet{2016GeoRL..43...50S} data-driven MHD simulation of this event with an analytic quantitative kinetic model for the radio emission concludes that the type II burst intervals of interest here are the fundamental emission.
To compare radio measurements with the CME propagation parameters,
we converted the radial distances yielded by the GCS fit $r_{\rm{GCS}}$ to frequencies $f$ using the density model of \citet{1999ApJ...523..812S} assuming the fundamental emission.
We consider the type II burst to be generated at both the CME leading edge ($f\sim r_{\rm{GCS}}$)
and flanks ($f \sim r_{\rm{GCS}}\cos{\lambda_{\rm{GCS}}}$).
The type II burst coincides well with the kinematic curve which corresponds to CME flanks (Figure \ref{fig:spectra}c).
Although type II bursts are generated at interplanetary shock fronts of propagating CMEs,
they predominantly appear close to CME flanks \citep[e.g.][]{1984A&A...134..222G,2014ApJ...791..115M}.
For our wave propagation analysis, we applied the singular value decomposition technique to multicomponent measurements
of auto- and cross-correlations of the voltages induced by electric field fluctuations \citep{2003RaSc...38a..10S,2012JGRA..11706101K}.
Figures \ref{fig:spectra}c --  \ref{fig:spectra}f present the calculated wave vector directions $\theta$ and $\phi$
in the radial-tangential-normal (RTN) coordinate system (a spacecraft centered coordinate system where the Sun is at ($\theta=90^\circ$, $\phi=0^\circ$), southward/northward is $\theta=0^\circ,180^\circ$,
and east/west is $\pm\phi=90^\circ$).
An intensity threshold ($S>5 \times 10^{-19}$~$\rm{W/m^2/Hz}$) has been applied to suppress the background.
The high frequency portions of the type II and type III bursts ($\sim>1$~MHz) propagate roughly towards \textit{STEREO-A},
whereas the low frequency portions propagate east of the spacecraft (Figure \ref{fig:spectra}e).
For \textit{STEREO-B}, the type II and type III bursts propagate east of the spacecraft
for over the entire interval analyzed (Figure \ref{fig:spectra}f).

At this time, the separation angle between the two STEREO spacecraft was 64$^\circ$ which allows us to accurately locate the sources
of the type II and type III bursts by triangulation
\citep{2014SoPh..289.3121K,2014SoPh..289.4633K}.
We note that this favorable separation angle would be also attained for combined spacecraft observations
from L$_1$ and the Fifth Sun-Earth Lagrangian point
\citep[L$_5$;][]{2011SPIE.8148E..0ZG,2011JASTP..73..658G,2015SpWea..13..197V}.
Data points around the maximum flux densities of the type II and type III bursts were carefully selected in order to perform the radio triangulation.
The triangulated type II and type III radio bursts are clustered along the CME propagation direction that is predicted by the GCS and SSEF fits,
with radial distances ranging from 0.12~au to 0.31~au (Figure \ref{fig:triang}).
We find that, for this interval, the radio sources are located considerably further from the Sun when compared to the density model of \citet{1999ApJ...523..812S},
which predicts radial distances between 0.03~au (1,500~kHz) and 0.10~au (300~kHz).
This ambiguity is probably caused by radio sources being spatially extended and/or radio propagation effects
\citep{1984A&A...140...39S,1985A&A...150..205S,2007ApJ...671..894T,2014SoPh..289.4633K}.
Averaged radio locations of the three analyzed intervals of the type II burst are as follows: $0.23\pm0.02$~au (22:30~UT),
$0.21\pm0.03$~au (00:30~UT), and $0.32\pm0.04$~UT (04:00~UT).
The source locations derived from the triangulation of the second interval are systematically located closer to the Sun than those of the first one,
when it would be expected to be further from the Sun.
This discrepancy is probably caused by the superposition of the signal of the second interval of the type II burst with a fast-drifting emission occurring at about the same time.
We thus will not include triangulated radio locations of the second interval in our analysis.
The calculated average speed of the CME-driven shock based on the triangulated source locations of the first and third interval is $656$~$\rm{kms^{-1}}$,
which is lower than those predicted by the GCS and SSEF fits.
The third interval was also observed by the Wind/Waves instrument at 332~kHz located at the L$_1$ point \citep{1995SSRv...71..231B}.
It yielded a wave vector azimuth $\phi_{\rm{RTN}}=-12^\circ$, which is a good agreement with the triangulated location by \textit{STEREO}/Waves,
which would be observed at $\phi_{\rm{RTN}}=-9^\circ$ by \textit{Wind}/Waves.

We have examined \textit{STEREO-A}/SECCHI/COR2 images in order to compare the relative positions
of the triangulated source locations of the first interval of the type II burst with coronal structures (Figure \ref{fig:secchi}c).
The type II burst sources are clustered along positions of two streamers located at the west limb and at the southern CME flank (see red arrows in Figure \ref{fig:secchi}c).
This suggests a close relationship between the type II burst and a possible interaction between the CME-driven shock and these streamers. 
It is consistent with the work of \citet{2014ApJ...791..115M} who reported a similar result for a different CME.
We note that another CME is already in progress along the south and appears to overlap with the radio sources. Inspection of the EUVI images shows that the CME is associated with a slow filament eruption starting earlier on the same day and appears to propagate at a plane about 90$^\circ$ away from the source of our CME. Therefore, it is unlikely that it is affecting our observations or interpretation. 

\subsection{In Situ Measurements}

Figure \ref{fig:plasma} shows measurements obtained by \textit{MESSENGER} and \textit{STEREO-A}. 
A fast forward interplanetary shock was observed by \textit{MESSENGER}, at 0.4~au, at around 14:30 UT on November 30 (Figure \ref{fig:plasma}a),
identified by a sharp increase in
the magnetic field magnitude (from 10 to 55~nT).
This shock was likely to have been driven by the CME analyzed in this study.
From the CME launch time and the arrival time of the shock at \textit{MESSENGER},
we estimate the average CME speed between the Sun and \textit{MESSENGER} to be $\sim$~$741$~$\rm{kms^{-1}}$.
The CME-driven shock was also detected in situ at \textit{STEREO-A}, at 0.96~au, at around 22:30 UT on December 1 (Figures \ref{fig:plasma}b -- \ref{fig:plasma}d).
Shock arrival corresponds to an abrupt increases in the magnetic field (from 7 to 20~nT),
proton density (1 to 12~cm$^{-3}$),
and proton bulk velocity (340 to 580~km~s$^{-1}$).
From the CME launch time and the time of shock arrival at \textit{STEREO-A}, we estimated the average CME speed to be 734~km~s$^{-1}$.
From the times of shock arrival at \textit{MESSENGER} and \textit{STEREO-A}, we estimate a CME speed between the two spacecraft of 729~km~s$^{-1}$.
The shock speed calculated by assuming the shock mass flux conservation equation over the shock, measured by \textit{STEREO-A}, is 602~km~s$^{-1}$.
Figure \ref{fig:plasma}e presents radio and plasma waves measurements recorded by the low frequency receiver on board \textit{STEREO-A} (2.5--160~kHz).
We observe large amplitude Langmuir waves and a harmonic emission of a slow drifting type II burst in upstream of the CME-driven shock.
We note that it is the first observation of the source region of an interplanetary type II burst by \textit{STEREO} \citep{2015JGRA..120.4126G}.

\subsection{Kinematics of the CME and CME-driven Shock}

Figure \ref{fig:orbit} shows an overview of the results obtained for the CME and CME-driven shock propagation.
We have averaged azimuths $\phi$ of the type II and type III burst source locations from which we derive their propagation directions.
Results for the propagation direction obtained for the type III burst are in a very good agreement with the results of GCS and SSEF fitting (Figure \ref{fig:orbit}a).
This suggest that in some cases type III bursts could be useful for estimating CME directions although they are related to solar flares instead of CMEs;
a more statistical approach is required to confirm whether this is the case generally.
Moreover, intense interplanetary type III bursts are more frequently observed than interplanetary type II bursts \citep{2011pre7.conf..325G}.
In contrast, the locations of the type II bursts seem to lie at the CME flanks. Hence they are consistent with the CME-driven shock rather than the driver itself. 
We note that the type III burst may not be related to this eruption since it occurred five hours after the CME liftoff.
However, it was the first complex type III burst observed since then, and we thus performed the radio triangulation
for a comparison with the CME kinematics.
Nonetheless, type II bursts occur sometimes without any type III bursts around.

Figure \ref{fig:orbit}b displays kinematics of the CME leading edge and radio sources.
The results from the various techniques are in good agreement with the CME-driven shock arrival times at \textit{MESSENGER} and \textit{STEREO-A}.
For an analysis of the type II frequency drift, we converted frequencies of the type II burst $f$ to radial distances $r$
using the density model of \citet{1999ApJ...523..812S}.
As in Figure~2, we consider two possible scenarios: the emission
originates at the CME leading edge ($r \sim f$) and flanks ($r/\cos{\lambda_{\rm{GCS}}} \sim f$).
A comparison with white-light and in situ measurements suggests that the latter is valid, and the derived average speed of the CME-driven shock from a frequency drift
is then $992\pm22$~km~s$^{-1}$, which is consistent with an assumption that the CME-driven shock propagates faster than the CME itself.
For further analysis of the type II burst, we use a relation between the drift rate $df/dt$ of radio emissions measured at the frequency $f$, and the radial source speed $v$:
\begin{equation}  \label{eq:man}
\frac{df}{dt}=\frac{fv}{2n}\frac{dn}{dr}\mbox{,}
\end{equation}
where $n$ is the density and $r$ is the radial distance \citep{1999A&A...348..614M}.
If we assume that below 1~MHz the CME driven shock is far enough that $n\sim r^{-2}$, we can calculate $\xi$ the angular deviation of the type II burst propagation direction
from radial by combining radio measurements with white-light observations as:
\begin{equation}  \label{eq:gopal}
\xi=\cos^{-1}\Bigg(\frac{fv_{\rm{WL}}}{\frac{df}{dt}r_{\rm{WL}}}\Bigg)\mbox{,}
\end{equation}
where $r_{\rm{WL}}$ and $v_{\rm{WL}}$ are the radial distance and the speed of the CME from white-light observations, respectively.
From radio and white-light parameters between the first and the second interval
($f=1.05$~MHz, $v_{\rm{WL(GCS)}}=761$~kms$^{-1}$, $df/dt=116$~Hzs$^{-1}$, and $r_{\rm{WL(GCS)}}=16$~R$_{\rm{S}}$),
we calculte $\xi$ to be $52^\circ$, whereas for parameters between the second and the third interval yield
$\xi=62^\circ$ ($f=450$~kHz, $v_{\rm{WL(SSEF)}}=862$~kms$^{-1}$, $df/dt=37$~Hzs$^{-1}$, and $r_{\rm{WL(SSEF)}}=32$~R$_{\rm{S}}$).
Our results confirm that the type II burst arises from the CME flanks
since the calculated values of $\xi$ are comparable to the CME angular half-width $\lambda_{\rm{GCS}}$.

\section{CONCLUSIONS}
\label{S-Conclusions}

\begin{figure*}[htb]
\centering
\includegraphics[angle=0,width=0.88\textwidth]{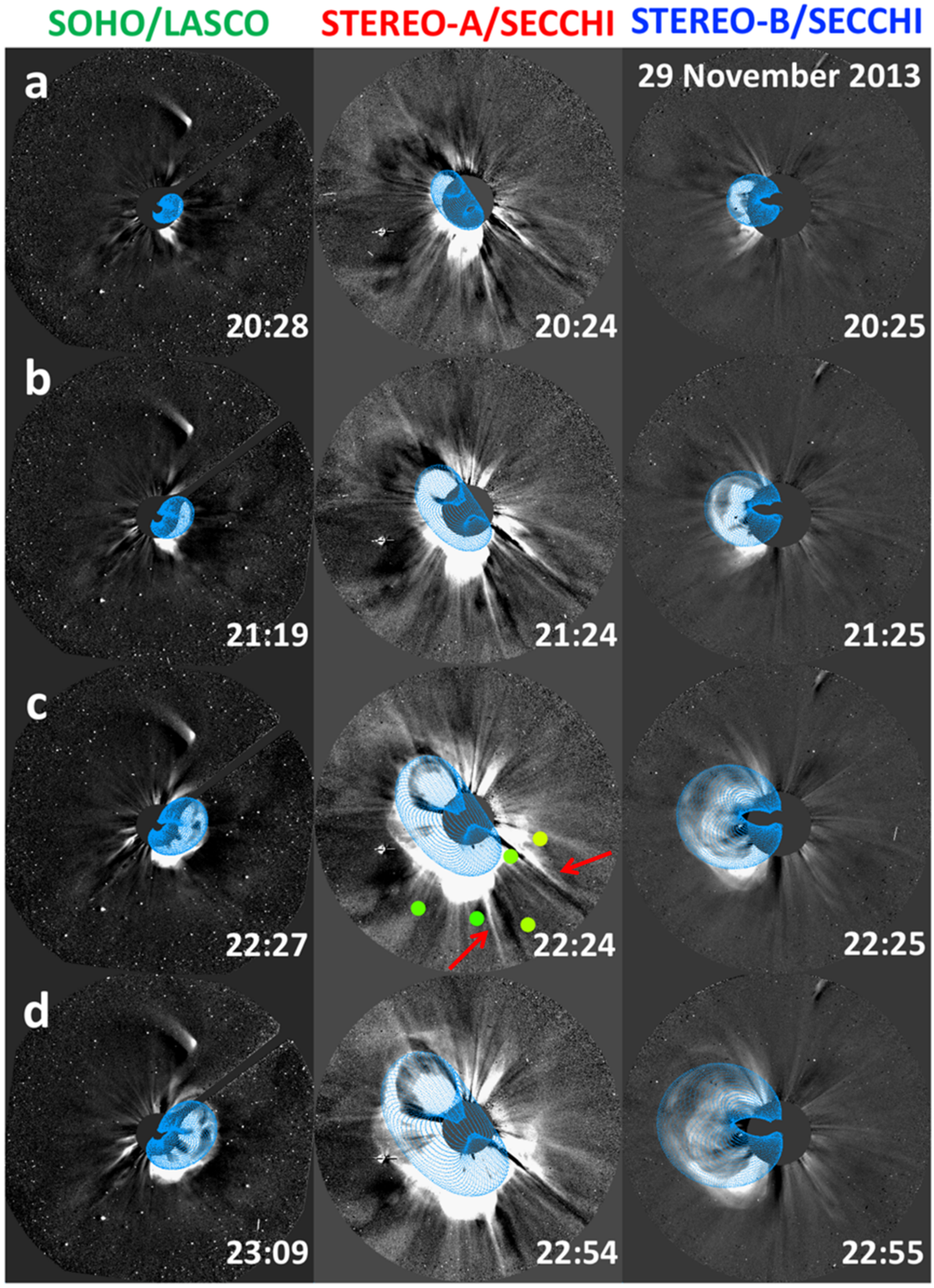}
\caption{White-light observations of the CME on 29 November 2013 by the \textit{SOHO}/LASCO/C3 and \textit{STEREO}/SECCHI/COR2 instruments.
(a -- d) Fit of the GCS model (blue grid) overlaid on multipoint coronagraph observations, from left to right: \textit{SOHO}, \textit{STEREO}-A, and \textit{STEREO}-B. 
Radio locations of the type II burst observed on 29 November are denoted by green circles in (c). Used colors correspond to those in Figure \ref{fig:triang}.}
\label{fig:secchi}
\end{figure*}

The CME that was launched from the Sun at around 16:00 UT on 2013 November 29 was well imaged by three widely-separated coronagraphs on board the \textit{STEREO-A}, \textit{STEREO-B}, and \textit{SOHO} spacecraft (Figure \ref{fig:secchi}). We were thus able to apply the GCS model to perform 3D reconstruction of the CME between $7$~$\rm{R_S}$ and $16$~$\rm{R_S}$.
We also applied the SSEF technique to the time-elongation profile of the CME apex extracted
from \textit{STEREO-A}/SECCHI/HI,
which allows us to track the CME from $30$~$\rm{R_S}$ to $140$~$\rm{R_S}$.
We find that the speeds and directions derived from GCS and SSEF are comparable. Furthermore, the CME-driven interplanetary shock was subsequently observed in situ at \textit{MESSENGER} and \textit{STEREO-A}, at $0.4$~au and  $1$~au from the Sun, respectively. The GCS and SSEF analysis can be used to make a reasonable prediction of
the arrival time of the CME-driven shock at \textit{MESSENGER} and \textit{STEREO-A}.

\begin{figure*}[htb]
\centering
\includegraphics[angle=0,width=0.88\textwidth]{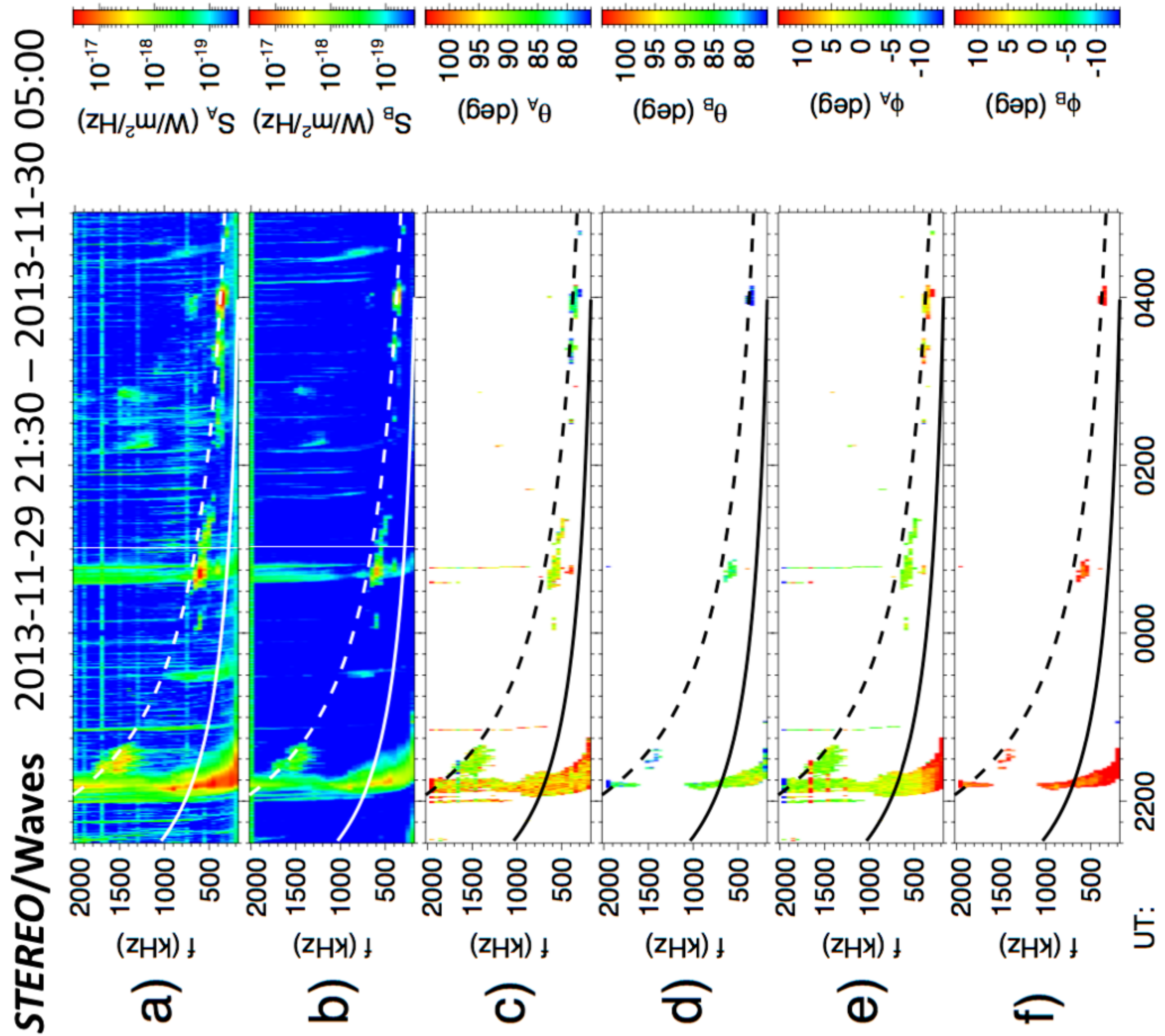}
\caption{Radio measurements of type II and type III bursts.
(a,b) Radio flux density $S$.
(c,d) Colatitude of wave vector $\theta$.
(e,f) Azimuth of wave vector $\phi$.
Solid and dashed lines denote results of the GCS model assuming emission at the leading edge and flanks, respectively.}
\label{fig:spectra}
\end{figure*}

\begin{figure}[htb]
\centering
\includegraphics[angle=0,width=0.45\textwidth]{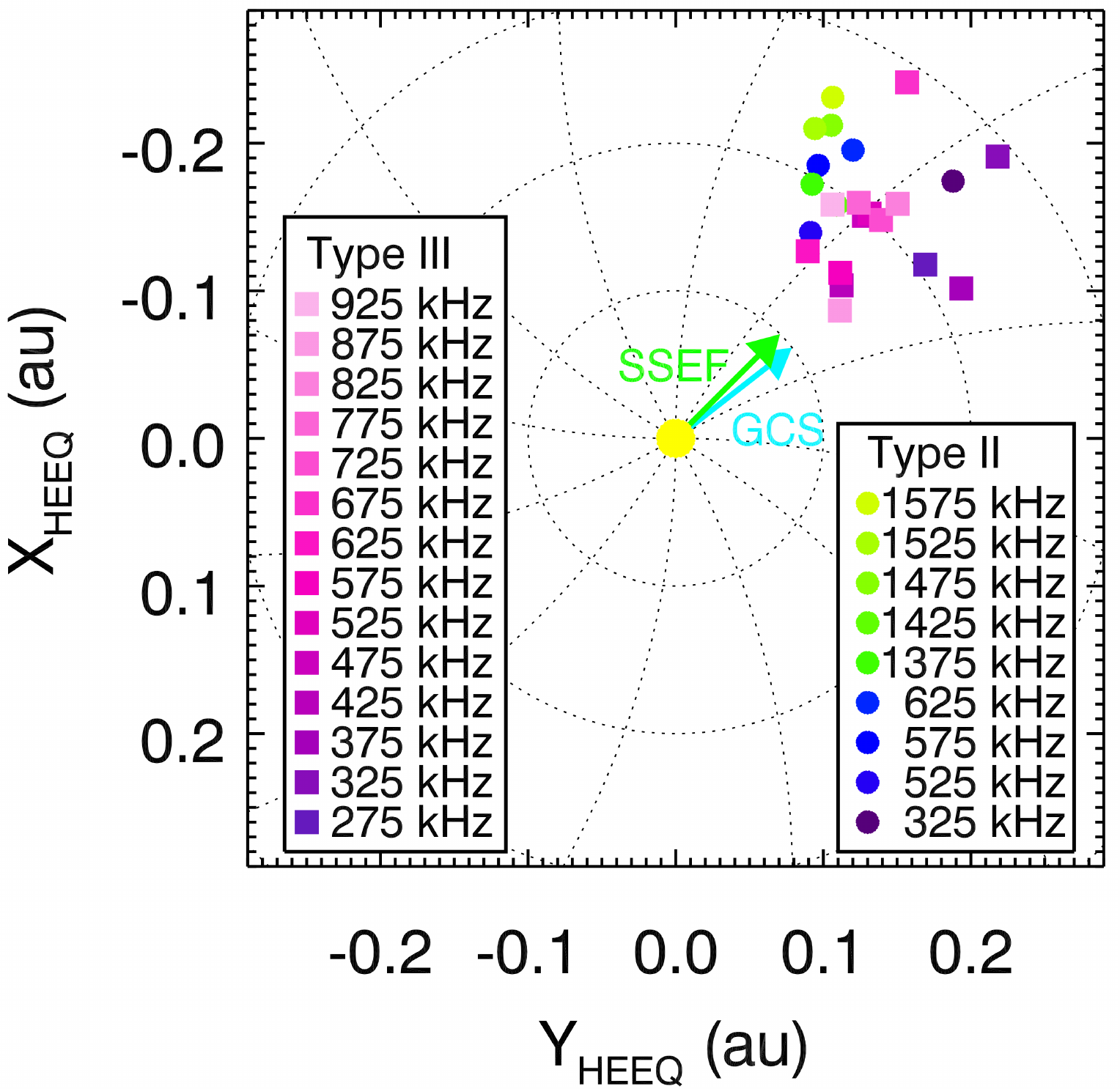}
\caption{Propagation analysis of radio measurements.
Radio source locations of type II (circles) and type III (squares) bursts for four time-frequency intervals in the XY$_{\rm{HEEQ}}$ plane. Colours denote frequencies. The cyan and green arrows indicate the CME propagation directions obtained by the GCS model
and the SSEF technique, respectively.
The Sun is at (0,0).}
\label{fig:triang}
\end{figure}

\begin{figure}[htb]
\centering
\includegraphics[angle=0,width=0.5\textwidth]{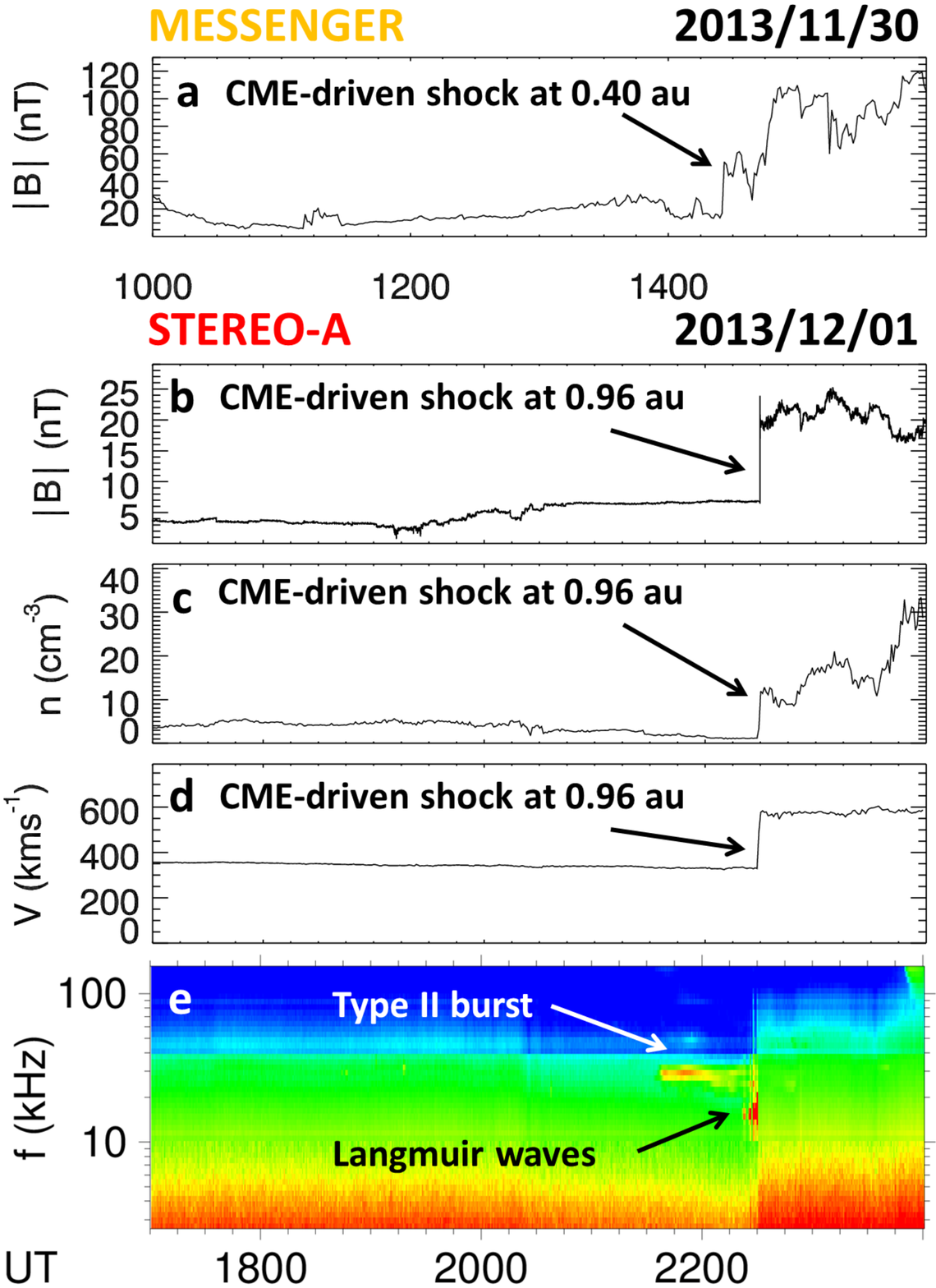}
\caption{Observations of the CME-driven shock at 0.40~au and 0.96~au.
(a) The magnetic field magnitude measured by \textit{MESSENGER} between 10:00 UT and 16:00 UT on 30 November 2013.
(b--e) The magnetic field magnitude, the proton density, the proton bulk speed, and electric field fluctuations
 recorded by \textit{STEREO}-A between 17:00 UT and midnight on 1 December 2013.}
\label{fig:plasma}
\end{figure}

\begin{figure}[htb]
\centering
\includegraphics[angle=0,width=0.45\textwidth]{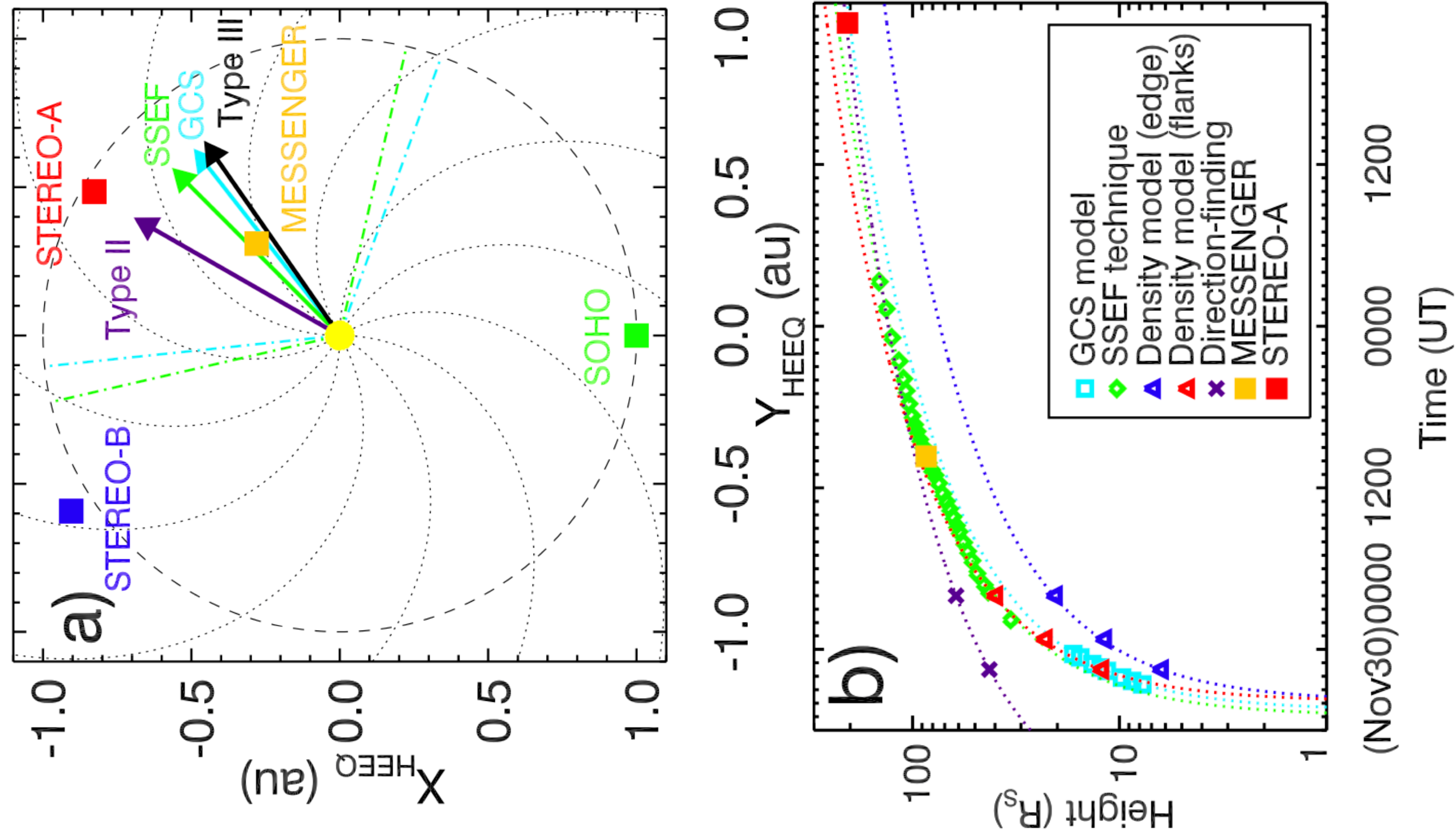}
\caption{a) Positions of the spacecraft in the solar equatorial plane on 29 November 2013.
The purple and arrows indicate average directions of type II and type III bursts, respectively.
The cyan and green arrows indicate the CME propagation directions obtained by the GCS model
and the SSEF technique, respectively. Dot-dashed lines show the CME half-width. b) Kinematics of the CME and radio sources between 2013 November 29 and December 1.
Dotted lines are linear fits.}
\label{fig:orbit}
\end{figure}

We have also performed an analysis of the type II and type III bursts associated with this event (Figure \ref{fig:spectra}).
We have successfully applied triangulation analysis to the three intervals of the type II burst and the single type III burst (Figure \ref{fig:triang}).
The relative positions of the first interval of the triangulated type II burst suggest that this emission is probably generated
by an interaction between the CME-driven shock and the two streamers (Figure \ref{fig:secchi}c).
The averaged directions, from triangulation, of the type III burst coincide radio sources with the CME direction from GCS and SSEF fitting,
while the directions of the type II burst indicate that their source regions are related to interaction of the CME-driven shock with the two streamers.

This CME would possibly trigger a geomagnetic storm if Earth-directed and lead to southward magnetic field at 1~au.
We thus conclude that interplanetary radio emissions
can provide us with an additional tool for predicting both CME-driven shock speed (using a frequency drift)
and direction (obtained by radio triangulation) with potential applications in space weather forecasting.
Currently, there are only two spacecraft in operation that carry coronagraphs (\textit{SOHO} and \textit{STEREO-A}) and
our ability to obtain reliable estimates of the time of arrival of CMEs and their shocks would be significantly curtailed if any one of them failed.  
We believe that the monitoring of CME-driven shocks, from both a scientific and space weather perspective,
would be enriched by the addition of space-born radio instrumentation.
In our view this would be best served by the monitoring interplanetary emissions from at least two vantage points,
preferably at L$_1$ and L$_5$.

\acknowledgements

The presented work has been supported by the European Union Seventh Framework Programme (FP7/2007--2013)
under grant agreement No. 606692 [HELCATS] and by the Praemium Academiae award of The Czech Academy of Sciences. 
O.~Kruparova, and J.~Soucek thanks support of the Czech Science Foundation grants GP16-16050Y, and GAP209/12/2394, respectively.
O.~Santolik acknowledges additional support from the LH15304 grant.
A.~Vourlidas was supported by NASA contract S-136361-Y to NRL.
V.~Bothmer acknowledges support of the CGAUSS (Coronagraphic German And US Solar Probe Plus Survey) project for WISPR by the German Space Agency DLR under grant 50 OL 1201.


\clearpage

\end{document}